\documentclass[a4]{article}

\font\scap=cmcsc10 \hfuzz=5cm
\font\scap=cmcsc10

\newcount\eqnumber
\def\neweq{{{(\the\eqnumber)}}\global\advance\eqnumber by 1}
\def\eqdef#1{\eqno\xdef#1{\the\eqnumber}\neweq}
\def\newaeq{{{(\the\eqnumber { a})}}\global\advance\eqnumber by 1}
\def\eqdaf#1{\eqno\xdef#1{\the\eqnumber}\newaeq}
\def\eqdisp#1{\xdef#1{\the\eqnumber}\neweq}
\def\eqdasp#1{\xdef#1{\the\eqnumber}\newaeq}

\newcount\refnumber
\def\newref{{\the\refnumber}\global\advance\refnumber by 1}
\def\refdef#1{{\xdef#1{\the\refnumber}}\newref}

\begin{document}

\centerline{\bf Miura transformations for discrete Painlev\'e equations coming from the affine E$_8$ Weyl group}
\bigskip
\medskip{\scap A. Ramani} and {\scap B. Grammaticos}
\quad{\sl IMNC, Universit\'e Paris VII \& XI, CNRS, UMR 8165, B\^at. 440, 91406 Orsay, France}

\medskip{\scap R. Willox}\quad
{\sl Graduate School of Mathematical Sciences, the University of Tokyo, 3-8-1 Komaba, Meguro-ku, 153-8914 Tokyo, Japan }

\bigskip
{\sl Abstract.}
\smallskip
We derive integrable equations starting from autonomous mappings with a general form inspired by the additive systems associated to the affine Weyl group E$_8^{(1)}$. By deautonomisation  we obtain two hitherto unknown systems, one of which turns out to be a linearisable one, and we show that both these systems arise from the deautonomisation of a non-QRT mapping. In order to unambiguously prove the integrability of these nonautonomous systems, we introduce a series of Miura transformations which allows us to prove that one of these systems is indeed a discrete Painlev\'e equation, related to the affine Weyl group E$_7^{(1)}$, and to cast it in canonical form. A similar sequence of Miura transformations allows us to effectively linearise the second system we obtain. An interesting off-shoot of our calculations is that the series of Miura transformations, when applied at the autonomous limit, allows one to transform a non-QRT invariant into a QRT one.

\bigskip
PACS numbers: 02.30.Ik, 05.45.Yv

\smallskip
Keywords: discrete Painlev\'e equations, Miura transformations, affine Weyl groups

\bigskip
1. {\scap Introduction}
\medskip

An important feature of Painlev\'e equations is the existence of a plenitude of transformations that establish relations between their various forms. This is particularly important in the case of discrete Painlev\'e equations, where there exist literally hundreds of different forms even if one only thinks of those that have been explicitly derived (the number of possible forms being infinite, as was shown in [\refdef\infinit]). The interrelations between discrete Painlev\'e equations can be of various types, involving Miura, B\"acklund [\refdef\gromak] or folding [\refdef\fold] transformations, to name but a few. This paper is devoted to transformations of Miura type. These transformations were named after the relation, discovered by R. Miura [\refdef\miura], that links the KdV and mKdV equations and which was the key to the proof of the integrability of these two systems.

In the domain of continuous Painlev\'e equations, the concept of a Miura transformation is best explained in the framework of the Hamiltonian representation of these equations, which was introduced by J. Malmquist [\refdef\malm] and F.J. Bureau [\refdef\bureau] and refined by K. Okamoto [\refdef\okamo]. According to this approach a Painlev\'e equation, say for the variable $x$, can be obtained from the Hamiltonian equations of motion
$${dx\over dt}={\partial H\over \partial p},\eqdef\one$$
$${dp\over dt}=-{\partial H\over \partial x},\eqdef\two$$
in terms of an explicitly time dependent Hamiltonian $H(x,p,t)$.
That is to say, starting from the Hamiltonian for some Painlev\'e equation and using the Hamiltonian equations of motion to eliminate the variable $p$, we obtain the corresponding Painlev\'e equation for $x$. On the other hand, when we eliminate the variable $x$ we will find another equation, for the variable $p$, which however also satisfies the Painlev\'e property (but which may be of higher degree in the second derivative [\refdef\cosgr]). The system of Hamiltonian equations thus constitutes a {\sl Miura transformation} connecting these two equations. The important remark here is that a Miura transformation for Painlev\'e equations should always be expressible as a system, allowing one to move from one equation to the other in a simple algebraic way, without the need to integrate any equations. 

Miura transformations also exist for discrete Painlev\'e equations, despite the absence of an explicit Hamiltonian formalism {\sl \`a la Okamoto}. The first such relation to be derived (to the authors' knowledge at least) is a Miura transformation [\refdef\ourmiura]  relating a discrete analogue of P$_{\rm II}$ to a discrete analogue of the equation known as P$_{34}$ (the number referring to the Gambier/Ince classification of ordinary differential equations that have the Painlev\'e property). This discrete Miura transformation is in perfect parallel to the one known in the continuous case. One starts by introducing the Miura system 
$$x_n=(y_n-1)(y_{n+1}+1)-z_n,\eqdef\three$$
$$y_n={m+x_{n-1}-x_n\over x_{n-1}+x_n},\eqdef\four$$
where $z_n$ is a given function of $n$, and $m$ is a free constant.
Eliminating the $x$ variable  one obtains for $y$ the d-P$_{\rm II}$ equation 
$$y_{n+1}+y_{n-1}={y_n(z_{n-1}+z_n)+m+z_n-z_{n-1}\over y_n^2-1},\eqdef\five$$
while the elimination of $y$ leads to
$$(x_{n+1}+x_n)(x_n+x_{n-1})={-4x_n^2+m^2\over x_n+z_n},\eqdef\six$$
which was shown in [\refdef\fokas] to be a discrete analogue of the P$_{34}$ equation.

The focus of this paper is on Miura transformations between equations that are obtained, through some limiting procedure, from systems associated with the affine Weyl group E$_8^{(1)}$ . In [\refdef\ourlimit] we derived discrete Painlev\'e equations associated with the groups E$_7^{(1)}$ and E$_6^{(1)}$, starting from an E$_8^{(1)}$ form. Here we shall revisit this question and show that there exist equations which are either associated to E$_7^{(1)}$ or are linearisable, but which at the autonomous limit are {\sl not} of QRT type [\refdef\qrt]  but rather belong to the class known as HKY, proposed initially by Hirota, Kimura and Yahagi [\refdef\hky]. Most importantly, the Miura transformations that we shall derive, when reduced to their autonomous form, show how one can transform an HKY mapping into a QRT one. Miura transformations in the case of  E$_6^{(1)}$-associated equations will also be presented. In all our derivations we concentrate on the additive case. Extending these results to the multiplicative case will be the object a future study.

\bigskip
2. {\scap From E$_8^{(1)}$ to E$_7^{(1)}$ and E$_6^{(1)}$}
\medskip

The general additive discrete Painlev\'e equation associated with the affine Weyl group $E_8^{(1)}$ has the form [\refdef\huit]: 
$$\displaylines{{(x_n-x_{n+1}+(z_n+z_{n+1})^2)(x_n-x_{n-1}+(z_n+z_{n-1})^2)+4x_n(z_n+z_{n+1})(z_n+z_{n-1})\over 
(z_n+z_{n-1})(x_n-x_{n+1}+(z_n+z_{n+1})^2)+(z_n+z_{n+1})(x_n-x_{n-1}+(z_n+z_{n-1})^2)}
\hfill\cr\hfill=2{x_n^4+S_2x_n^3+S_4x_n^2+S_6x_n+S_8\over S_1x_n^3+S_3x_n^2+S_5x_n+S_7},\quad\eqdisp\seven\cr}$$
where $z_n$ is equal to $\alpha n+\beta$.
The coefficients $S_k$ are elementary symmetric functions of the quantities $z_n+\kappa_n^i$, where the $\kappa^i_n$ are eight parameters which are, generically, functions of the independent variable. However, a more convenient form was proposed in [\refdef\ourdegen]. It is obtained through the introduction of an ancillary variable $\xi$ which is just the square root of $x$. Thus, with $x_n=\xi_n^2$, we find that (\seven) can be rewritten as
$${x_{n+1}-(\xi_n-z_n-z_{n+1})^2\over x_{n+1}-(\xi_n+z_n+z_{n+1})^2}\,{x_{n-1}-(\xi_n-z_n-z_{n-1})^2\over x_{n-1}-(\xi_n+z_n+z_{n-1})^2}={\prod_{i=1}^8(\kappa_n^i+z_n-\xi_n)\over\prod_{i=1}^8(\kappa_n^i+z_n+\xi_n)}.\eqdef\eight$$
This form (also obtained by Kajiwara, Noumi and Yamada in [\refdef\kanoya]) is particularly convenient for the calculations necessary for singularity analysis.

In order to go from this general $E_8^{(1)}$-type equation to equations associated with the E$_7^{(1)}$ and E$_6^{(1)}$ affine Weyl groups, one has to implement specific limits. The simplest limit is obtained by assuming that two of the parameters $\kappa$ go to infinity while their sum remains finite. This results in an equation of the form (\eight) where the right-hand side is a ratio of 6 (instead of 8) products or, equivalently, in an equation of the form (\seven) where the right-hand side is a ratio of a cubic over a quadratic polynomial. As shown in [\ourlimit], this is the general $E_7^{(1)}$-type equation. Taking a second limit where yet another pair of $\kappa$ tend to infinity yields linearisable mappings. There is however a different way in which a limit can be obtained: we can let three of the parameters $\kappa$ tend to infinity, while ensuring that their sum remains finite. In this case we find an equation of the form (\eight) with a right-hand side that is a ratio of 5 products or, equivalently, an equation of the form (\seven) with a ratio of quadratic polynomials as its right-hand side. These limits were also examined in [\ourlimit] where we showed that they result in the general $E_6^{(1)}$-type equation. Here again, letting two more parameters $\kappa$ tend to infinity will lead to linearisable equations. 

Once the general equation of a certain type is obtained one can proceed further by considering possible simplifications in the right-hand side. All these simplifications lead to discrete Painlev\'e equations associated to the same affine Weyl group as the initial one, and have been studied in detail in [\ourdegen] and [\ourlimit]. Implementing all possible simplifications in the case of $E_7^{(1)}$-type equations (in the form (\seven)) one arrives at a right-hand side which is simply linear in $x$, while in the case of $E_6^{(1)}$-type equations the resulting right-hand side is of degree zero in $x$ i.e., it is simply a combination of the functions $z_n$. 

However this is not the only possibility. There might exist mappings of similar form which do not arise as limits and/or simplifications of an $E_8^{(1)}$ equation.
At this point an interesting experiment suggests itself. Suppose we consider an equation of the form (\seven) where the right-hand side is independent of $x_n$. Is it possible to recover the results obtained in [\ourlimit] concerning the $E_6^{(1)}$-type equations, simply relying on algebraic entropy techniques?

We shall perform this calculation in the autonomous case, as this is the usual starting point for systems that are obtained by simplification from `higher' ones. We shall therefore consider a mapping of the form
$${(x_n-x_{n+1}+4z^2)(x_n-x_{n-1}+4z^2)+16x_nz^2\over 2z(x_n-x_{n+1}+4z^2)+2z(x_n-x_{n-1}+4z^2)}=Nz,\eqdef\nine$$
for some integer $N$ and constant $z$.

We start from initial conditions $x_0$ and $x_1=p/q$ and we calculate the homogeneous degree in $p,q$ of the successive iterates $x_n$. All odd values of $N$, as well as all $N\le0$ lead to an exponential degree growth, which means that all these cases are non-integrable. Performing the same calculation for even $N>0$, we first of all find for $N=2$ that the degrees saturate as 0,1,2,2,2,$\cdots$, which means that the mapping is linearisable. This is indeed the case (the details of the linearisation are  given in [\ourlimit], equation (42)). For $N=4$ we find the succession of degrees 0,1,2,3,4,4,$\cdots$, which is again an indication of linearisability. However the corresponding mapping does not appear to be known and in section 4 we shall therefore proceed to its deautonomisation and linearisation. For $N=6$ we find for the degrees 0,1,2,3,5,7,9,11,13,15,17,19,21,23,25,27,29,$\cdots$, i.e. again a linear growth. The corresponding mapping was already derived in [\ourlimit], equation (38), where it was linearised after deautonomisation. 

In the case $N=8$ the degrees growth clearly quadratically: 0,1,2,3,5,7,10,14,18,22,27,33,39,46,53,$\cdots$, which is an indication that the mapping should be integrable and not  linearisable. This is also a new result, at least to the authors' knowledge. The integrability of the autonomous mapping is established by the existence of the invariant
$$K={(x_n+x_{n-1}-20z^2)(x_n-4z^2)(x_{n-1}-4z^2)((x_n-x_{n-1})^2-16z^4)\over((x_n+x_{n-1}-4z^2)^2-4x_nx_{n-1})^2},\eqdef\ten$$
which is conserved by the mapping (\nine) when $N=8$. It is important to note that (\ten) is quartic in $x_n$ and $x_{n-1}$ and thus does not belong to the QRT family, but is of HKY type. The deautonomisation of this $N=8$ case will be presented in the next section, where it will be shown that it actually leads to a Painlev\'e equation associated to the affine Weyl group E$_7^{(1)}$.

Finally, for $N=10$ we again find quadratic degree growth 0,1,2,2,3,5,7,9,11,14,18,21,24,29,34,$\cdots$, in agreement with the results of [\ourlimit] where we showed that the corresponding non-autonomous form (equation (27) under constraint (34) in [\ourlimit]) is in fact a discrete Painlev\'e equation associated with the affine Weyl group $E_6^{(1)}$. For even $N$, beyond the value of 10, all cases examined led to exponential growth. Thus we surmise that the only integrable cases that exist are the five cases identified here, two of which appear to be new.

\bigskip
3. {\scap Deautonomisation and Miura transformations for the E$_7^{(1)}$ system}
\medskip

Before proceeding to the deautonomisation of the mapping (\nine) with $N=8$, we must first perform its singularity analysis [\refdef\sincon]. We readily find that a singularity appears when $x_n$ takes the value $16z^2$. Iterating further, we find the succession of values $4z^2, 0,4z^2, 16z^2$ and a value for $x_{n+5}$ that again depends on $x_{n-1}$. Thus we conclude that the singularity is confined and we can proceed to the actual deautonomisation of the mapping by requiring that its nonautonomous version has a singularity pattern of the same form and length as the autonomous one. We shall not go into all the arithmetic detail of the actual deautonomisation of the mapping. Suffice it to say that in order to obtain a similar singularity pattern, we must introduce an auxiliary function $\zeta_n$ such that $z_n+z_{n-1}=\zeta_{n+2}+\zeta_{n-2}$, and replace the ``$8z$'' contribution in the right-hand side by the quantity $2(\zeta_{n+2}+\zeta_{n+1}+\zeta_{n}+\zeta_{n-1})$.  We thus find the equation
$$\displaylines{{(x_n-x_{n+1}+(z_n+z_{n+1})^2)(x_n-x_{n-1}+(z_n+z_{n-1})^2)+4x_n(z_n+z_{n+1})(z_n+z_{n-1})\over 
(z_n+z_{n-1})(x_n-x_{n+1}+(z_n+z_{n+1})^2)+(z_n+z_{n+1})(x_n-x_{n-1}+(z_n+z_{n-1})^2)}
\hfill\cr\hfill=2(\zeta_{n+2}+\zeta_{n+1}+\zeta_{n}+\zeta_{n-1}).\quad\eqdisp\done\cr}$$
With this re-parametrisation, the singularity that appears when $x_n=(\zeta_{n+2}+\zeta_{n+1}+\zeta_{n}+\zeta_{n-1})^2$ is automatically confined, without any further constraints on $\zeta_n$. However, this mapping has another singularity when $x_n$ goes to infinity. Starting from a regular value for $x_{n-1}$ and $x_n=\infty$, we obtain three more infinities and a possibility for confinement at the level of $x_{n+4}$. Performing the calculation in detail we find that confinement is indeed possible, provided that $\zeta_n$ satisfies the constraint 
$$\zeta_{n+6}-\zeta_{n+3}-\zeta_{n+1}+\zeta_{n-2}=0.\eqdef\dtwo$$
The solution to (\dtwo) is $\zeta_n=\alpha n+\beta+\phi_3(n)+\phi_5(n)$, where $\phi_m(n)$ is a periodic function with period $m$, i.e. $\phi_m(n+m)=\phi_m(n)$, given by
$$ \phi_m(n)=\sum_{\ell=1}^{m-1} \epsilon_{\ell}^{(m)} \exp\left({2i\pi \ell n\over m}\right),\eqdef\dthree$$
in terms of $m-1$ free (constant) parameters $\epsilon_\ell^{(m)}$.
Notice that the summation in (\dthree) starts at 1 instead of 0 and that $\phi_m(n)$ therefore does not contain a constant term (the presence of which would have led to awkward notations when more than one such $\phi$ appears). Thus, since any $\phi_m$ introduces $(m-1)$ parameters, we find that in the present case the number of parameters in (\done) is 7 (excluding $\alpha$, which is just the size of the discretisation step) and we therefore expect this equation to be associated to the affine Weyl group $E_7^{(1)}$. As we shall show in the following, this is indeed the case.

Having obtained the non-autonomous form (\done) of (\nine) for $N=8$, we can now proceed to the construction of its  Miura transformations which will, eventually, allow us to identify this mapping as an $E_7^{(1)}$-type mapping. We first introduce a new variable, $w_n$, obtained from $x_n$ through the relation
$$w_n={x_n-x_{n-1}\over 2(\zeta_{n+2}+\zeta_{n-2})}-{1\over2}(\zeta_{n+2}-\zeta_{n-2})-\zeta_{n+1}+\zeta_{n-1}.\eqdef\dfour$$
This is only the first half of a Miura transformation, and this relation should be complemented with 
$$\displaylines{{x_n=w_{n+1}w_n-w_{n+1}(\zeta_n+2\zeta_{n-1})+w_n(\zeta_{n+1}+2\zeta_{n+2})+(\zeta_n+\zeta_{n-1})^2+(\zeta_{n+1}+\zeta_{n+2})^2}
\hfill\cr\hfill+\zeta_n\zeta_{n+1}-2\zeta_{n-1}\zeta_{n+2}.\quad\eqdisp\dfive\cr}$$
Eliminating $x$ between (\dfour) and (\dfive) leads to the following equation for the variable $w$:
$$w_{n+1}(\zeta_n+2\zeta_{n-1}-w_n)+w_{n-1}(\zeta_n+2\zeta_{n+1}+w_n)-w_n(\zeta_{n+1}+\zeta_{n-1})+\zeta_n(\zeta_{n+1}-\zeta_{n-1})=0.\eqdef\dsix$$
This is also an equation which, at the autonomous limit, does not lead to a QRT mapping since it has the invariant
$$K={(w_n^2-\zeta^2)(w_{n-1}^2-\zeta^2)(w_nw_{n-1}-3\zeta w_n+3\zeta w_{n-1}+3\zeta^2)\over (w_n-w_{n-1}-2\zeta)^2},\eqdef\dseven$$
where we have taken $\zeta_n\equiv\zeta$. We remark that this invariant is of HKY-type since its numerator is cubic in $w_n$ and $w_{n-1}$, but most importantly, that this degree is lower than that of the invariant  (\ten) for the mapping before the Miura transformation. 

A second Miura is therefore in order. We introduce a new variable $y_n$, related to $w_n$ through
$$y_n={(w_{n+1}+\zeta_{n+1})(\zeta_n-w_n)\over2(w_{n+1}-w_n-\zeta_{n+1}-\zeta_n)},\eqdef\deight$$
$$w_n=\zeta_n{y_{n-1}-y_n\over y_{n-1}+y_n},\eqdef\dnine$$
which leads to the equation
$$y_{n+1}y_{n-1}(\zeta_{n+1}+\zeta_n)+y_{n}y_{n-1}\zeta_{n}+y_{n+1}y_{n}\zeta_{n+1}+y_{n}\zeta_n\zeta_{n+1}=0.\eqdef\dten$$
We have now reached an equation which, at the autonomous limit, does indeed become a mapping of QRT type. Its invariant is
$$K={y_ny_{n-1}(y_n+\zeta)(y_{n-1}+\zeta)\over(y_n+y_{n-1})(y_n+y_{n-1}-\zeta)},\eqdef\vone$$
i.e. a ratio of polynomials that are quadratic in $y_n$ and $y_{n-1}$. Thus we conclude that, starting from the HKY mapping for $x$ with bi-quartic invariant (\ten), by combining the Miura transformations (\dfour-\dfive) and (\deight-\dnine), we can transform it into a mapping for $y$ with an invariant of QRT type.
 
 Equation (\dten) can be cast into the canonical form of equations associated to the $E_7^{(1)}$ affine Weyl group. We readily find 
 $$\left({y_n+y_{n+1}-\zeta_{n+1}\over y_n+y_{n+1}}\right)\left({y_n+y_{n-1}-\zeta_{n}\over y_n+y_{n-1}}\right)={y_n-\zeta_{n+1}-\zeta_n\over y_n},\eqdef\vtwo$$
 which is a discrete Painlev\'e equation already identified in [\refdef\seven].
 
Let us now briefly proceed in a different direction and examine the equation that can be obtained for $w$ if we consider its evolution along indices of the same parity, i.e. if we consider an equation relating $w_n$ to $w_{n\pm2}$. What one obtains is a very lengthy equation, which we shall not give here (its derivation being elementary). However, it is worth pointing out that if, to this equation, we apply a new Miura transformation that involves the variable $u_n$, introduced through
$$\displaylines{\Big(w_{n-1}-w_{n+1}+\zeta_{n+1}+2\zeta_{n}+\zeta_{n-1}\Big)u_n=2w_{n-1}w_{n+1}(\zeta_{n+1}+\zeta_{n}+\zeta_{n-1})+w_{n-1}(\zeta_{n-1}^2-\zeta_{n+1}^2-2\zeta_{n}\zeta_{n+1})\hfill\cr\hfill
+w_{n+1}(\zeta_{n-1}^2-\zeta_{n+1}^2+2\zeta_{n}\zeta_{n-1})+(\zeta_{n+1}+2\zeta_n+\zeta_{n-1})(\zeta_{n+1}^2+\zeta_{n-1}^2)+2\zeta_n\zeta_{n+1}\zeta_{n-1},
\quad\eqdisp\vthree\cr}$$
complemented with
$$w_n=-{u_{n+1}-u_{n-1}\over2(\zeta_{n+2}+\zeta_{n-2})}+{1\over2}(\zeta_{n+2}-\zeta_{n-2}),\eqdef\vfour$$
we obtain for $u_n$ the (much more manageable) equation  
$$\displaylines{{(u_n-u_{n+2}+(\zeta_{n-1}+\zeta_{n+3})^2)(u_n-u_{n-2}+(\zeta_{n+1}+\zeta_{n-3})^2)+4u_n(\zeta_{n-1}+\zeta_{n+3})(\zeta_{n+1}+\zeta_{n-3})\over (\zeta_{n+1}+\zeta_{n-3})(u_n-u_{n+2}+(\zeta_{n-1}+\zeta_{n+3})^2)+(\zeta_{n-1}+\zeta_{n+3})(u_n-u_{n-2}+(\zeta_{n+1}+\zeta_{n-3})^2)}\hfill\cr\hfill={u_n+(\zeta_{n+1}+2\zeta_{n}+\zeta_{n-1})(\zeta_{n+1}+\zeta_{n-1})\over\zeta_{n+1}+\zeta_{n}+\zeta_{n-1}}.\quad\eqdisp\vfive\cr}$$
In fact this equation was already identified in [\ourlimit].
Combining the Miura transformations (\dfour-\dfive) and (\vthree-\vfour) it is possible to write a Miura transformation that links the variable $x$ directly to the variable $u$. Its expression is too long to be given here but it is of course perfectly amenable to calculations with the help of computer algebra. 

Thus, thanks to the chain of Miura transformations (\dfour-\dfive) and (\deight-\dnine), we were able to bring equation (\done), which we expected to be associated with the group E$_7^{(1)}$, to a form which is the canonical one for equations of the group E$_7^{(1)}$. Similarly, the Miura transformations (\vthree)-(\vfour) allow us to bring the initial equation, {\sl when we consider one point out of two}, to a form that is indeed canonical for equations obtained from a system related to the group E$_8^{(1)}$, through the appropriate limit.

\bigskip
4. {\scap The linearisable equation}
\medskip

As explained in section 2 our exploration of equation (\nine) resulted in two new integrable systems, corresponding to the cases $N=8$ and $N=4$, the first one of which was analysed in the previous section. Here we shall deal with the $N=4$ case. 

In order to deautonomise this mapping we start from the ansatz
$$\displaylines{{
{(x_n-x_{n+1}+(z_n+z_{n+1})^2)(x_n-x_{n-1}+(z_n+z_{n-1})^2)+4x_n(z_n+z_{n+1})(z_n+z_{n-1})\over 
(z_n+z_{n-1})(x_n-x_{n+1}+(z_n+z_{n+1})^2)+(z_n+z_{n+1})(x_n-x_{n-1}+(z_n+z_{n-1})^2)}}
\hfill\cr\hfill=2Z_{n}+2Z_{n+1},\quad\eqdisp\vsix\cr}$$
where $z_n$ and $Z_n$ are two functions to be determined. To do so we compute the degree growth of (\vsix) and require that the successive degrees be the same as in the autonomous case. Iterating the mapping for the same initial conditions as in section 2, we find the degrees 0,1,2,3, and requiring that the degree of $x_n$ be 4 for all $n\ge4$,  we obtain the condition
$$Z_{n+1}+Z_{n-1}=z_{n-1}+z_n. \eqdef\vseven$$
This condition can be integrated by introducing an auxiliary variable $\zeta_n$, resulting in:
$$Z_n=\zeta_{n-1}+\zeta_n\quad{\rm and}\quad z_n=\zeta_{n+1}+\zeta_{n-1}.\eqdef\veight$$
In order to integrate (\vsix) -- which now contains a single free function, $\zeta_n$ -- we start by introducing a first Miura transformation, involving a new variable $w_n$. With hindsight we define
$$w_n={x_n-x_{n-1}\over2(Z_{n+1}+Z_{n-1})}+{Z_{n-1}-Z_{n+1}\over 2}, \eqdef\vnine$$
complemented with
$$x_n=(w_{n+1}+Z_{n+1})(w_n-Z_n)+(Z_n+Z_{n+1})^2.\eqdef\vten$$
We readily find the following linear equation for $w_n$:
$$w_{n+1}(Z_{n}-w_n)+w_{n-1}(Z_n+w_n)+w_n(Z_{n+1}+Z_{n-1})+Z_n(Z_{n-1}-Z_{n+1})=0.\eqdef\tone$$
If one defines the quantity $k_n$
$$k_n=(w_{n-1}+Z_{n-1})/(w_n-Z_n),\eqdef\ttwo$$ 
then equation (\tone) can be rewritten as 
$$k_n k_{n+1}=1,\eqdef\tthree$$ 
and  thus $k_n=c^{(-1)^n}$ for some constant $c$. This allows us to integrate (\tone), whereby also obtaining the solution for (\vsix) through (\vten).
One can, in fact, give an explicit solution for $w_n$. From (\tone) one can check that
$$ {w_{n+1}-w_{n-1}-Z_{n+1}-Z_{n-1}\over 2Z_n}=k_n=c^{(-1)^n}.\eqdef\tfour$$
From (\tfour) one can write a linear equation for $w$ of a given parity. 
For instance, for $w$ of odd parity one obtains
$$w_{2m+1}-w_{2m-1}=Z_{2m+1}+Z_{2m-1}+2c Z_{2m}.\eqdef\tfive$$
Let us introduce the semi-infinite sum $S_n=\sum^n\zeta(j)$, defined up to an additive contant which is the other integration constant besides $c$. One obviously has $S_n-S_{n-2}=\zeta_n+\zeta_{n-1}=Z_n$. Hence 
$$w_{2m+1}=S_{2m+1}+S_{2m-1}+2c S_{2m},\eqdef\tsix$$
where the integration constant is ``rolled into'' the definition of $S_n$. Solving (\tone) for $w_{2m}$ we find 
$w_{2m}=S_{2m}+S_{2m-2}+2c^{-1} S_{2m-1}.$
with the {\sl same} integration constant for $S_m$ as before. This allows us to write  the general solution as 
$$w_{n}=S_{n}+S_{n-2}+2c^{(-1)^{n-1}} S_{n-1}.$$

At this point it is interesting to go back to the autonomous case, assuming that $z_n$ and $Z_n$ take the same constant value $z$ for all $n$. Since the quantity $k_n$ is inverted at each step it follows that the conserved quantity $K$ is just $k_n+k_n^{-1}$. In terms of ($w_n,w_{n-1}$) we obtain 
$$K(w_n,w_{n-1})={w_n^2+w_{n-1}^2-2zw_n+2zw_{n-1}+2z^2\over(w_n-z)(w_{n-1}+z)}.\eqdef\tseven$$
We can easily verify that the evolution of the autonomous limit of (\tone) preserves the conservation $K(w_n,w_{n-1})=K(w_{n+1},w_n)$. Using (\vnine) we can of course obtain
$$8z^2(K(x_n,x_{n-1})-2)={((x_n+x_{n-1}-4z^2)^2-4x_nx_{n-1})^2\over((x_n-x_{n-1})^2-16z^4)(x_n+x_{n-1}-4z^2)},\eqdef\teight$$
The evolution of (the autonomous limit of) equation (\vsix) preserves the conservation law $K(x_n,x_{n-1})=K(x_n,x_{n+1})\, (\equiv K(x_{n+1},x_n)$). The constant 2 was subtracted to allow the numerator to factorise to a perfect square, which is in fact that of the denominator of the standard invariant for the autonomous limit of additive type equations related to the E$_8^{(1)}$ affine Weyl group. It is important to point out that the invariant (\teight) is not of QRT, but rather of HKY type since the variables $x_n$ and $x_{n-1}$ enter with powers higher than 2.

As $k_n$ takes a very simple form when expressed in terms of $w_{n+1}$ and $w_{n-1}$, $k_n= {(w_{n+1}-w_{n-1}-2z)/2z}$, we introduce the variable $y_n=w_{n+1}-w_n-z$ so that $w_{n+1}-w_{n-1}-2z=y_{n}+y_{n-1}$. The invariant $K=k_n+k_n^{-1}$ can then be written as 
$$K(y_n,y_{n-1})={(y_n+y_{n-1})^2+4z^2\over 2z(y_n+y_{n-1})}.\eqdef\tnine$$
This invariant is clearly of QRT form. The transformation from $w_n$ to $y_n$, being invertible through $w_n=z(y_n-y_{n-1})/(y_n+y_{n-1}-2z)$, is also a Miura transformation. So here again, thanks to the chain of Miura transformations we introduced, we were able to bring a mapping of HKY type to a QRT form, namely $(y_{n+1}+y_n)(y_n+y_{n-1})=4z^2$.

\bigskip
5. {\scap Deautonomisation and Miura transformations for an E$_6^{(1)}$ system} 
\medskip

Having obtained quite interesting Miura transformations for the new E$_7^{(1)}$-type discrete Painlev\'e equation in the case $N=8$, one could wonder whether such transformations also exist when $N=10$, which we know to yield a discrete Painlev\'e equation associated to the affine Weyl group  E$_6^{(1)}$ [\ourlimit]. It turns out that this is indeed the case.

In [\ourlimit], the deautonomisation of the ``$10z$'' equation was shown to lead to the equation 
$$\displaylines{{{(x_n-x_{n+1}+(z_n+z_{n+1})^2)(x_n-x_{n-1}+(z_n+z_{n-1})^2)+4x_n(z_n+z_{n+1})(z_n+z_{n-1})\over 
(z_n+z_{n-1})(x_n-x_{n+1}+(z_n+z_{n+1})^2)+(z_n+z_{n+1})(x_n-x_{n-1}+(z_n+z_{n-1})^2)}}
\hfill\cr\hfill=4z_{n+1}+2z_{n}+4z_{n-1},\quad\eqdisp\tten\cr}$$
where $z_n=\alpha n+\beta+\gamma(-1)^n+\phi_3(n)+\chi_4(n)$, and where the (anti-)periodic function $\chi_{2m}$ obeys the equation $\chi_{2m}(n+m)+\chi_{2m}(n)=0$. It therefore has period $2m$, while involving only $m$ parameters, and can be expressed in terms of roots of unity as
$$ \chi_{2m}(n)=\sum_{\ell=1}^{m} \eta_{\ell}^{(m)} \exp\left({i\pi(2\ell-1)n\over m}\right).\eqdef\qone$$
The equation (\tten) thus depends only on six free parameters, showing that is indeed to be associated with the affine Weyl group  E$_6^{(1)}$.

Next, we introduce the Miura transformation defined by the system
$$w_n={x_n-x_{n-1}\over2(z_n+z_{n-1})},\eqdef\qtwo$$
$$\displaylines{4x_n=4w_nw_{n+1}+2w_n(z_n+4z_{n-1}+3z_{n+1})-2w_{n+1}(z_n+3z_{n-1}+4z_{n+1})\hfill\cr\hfill+3z_n^2+9z_n(z_{n-1}+z_{n+1})+4z_{n-1}^2+7z_{n-1}z_{n+1}+4z_{n+1}^2.\quad\eqdisp\qthree\cr}$$
Eliminating $x$ between equations (\qtwo) and (\qthree) we obtain, for $w_n$, an equation which can be written in a nice factorised form as
$${w_n+w_{n+1}-z_{n+1}/2+2z_n+z_{n-1}/2+4z_{n-2}\over w_n+w_{n-1}+z_{n-2}/2-2z_{n-1}-z_n/2-4z_{n+1}}={w_n+3z_n/2+z_{n-1}/2+2z_{n-2}\over w_n-3z_{n-1}/2-z_n/2-2z_{n+1}}.\eqdef\qfour$$
At this point it is convenient to introduce another function $\zeta_n$ closely related to $z_n$, the precise expression of which is $\zeta_n=\alpha n+\beta-3\gamma(-1)^n+\phi_3(n)-\chi_4(n)$ or equivalently $\zeta_n=z_{n+3}+z_{n-3}-z_n$. 
Moreover we introduce the new dependent variable $u_n$ as a translation of $w_n$:
$$u_n=w_n+{1\over2}(3z_n+z_{n-1}+4z_{n-2})-\zeta_n-\zeta_{n-1}-2\zeta_{n+1}.\eqdef\qfive$$
This allows us to write equation (\qfour) in a slightly simpler form as
$${u_n+u_{n+1}+\zeta_{n-1}+2\zeta_n+3\zeta_{n+1}\over u_n+u_{n-1}-\zeta_n-2\zeta_{n-1}-3\zeta_{n-2}}={u_n+\zeta_{n-1}+\zeta_n+2\zeta_{n+1}\over u_n-\zeta_n-\zeta_{n-1}-2\zeta_{n-2}}.\eqdef\qsix$$
We can now introduce a second Miura transformation:
$$y_n={(u_n-\zeta_n-\zeta_{n-1})(u_{n+1}+\zeta_n+\zeta_{n+1})\over2(u_n-u_{n+1}+\zeta_{n-1}+\zeta_{n+1})},\eqdef\qseven$$
$$u_n=(\zeta_n+\zeta_{n-1}){y_{n-1}-y_n\over y_{n-1}+y_n}.\eqdef\qeight$$
This leads to an equation for $y$ of the form
$$\left({y_n+y_{n+1}-\zeta_n-\zeta_{n+1}\over y_n+y_{n+1}}\right)\left({y_n+y_{n-1}-\zeta_n-\zeta_{n-1}\over y_n+y_{n-1}}\right)={y_n-\zeta_{n+1}-\zeta_n-\zeta_{n-1}\over y_n}.\eqdef\qnine$$
We remark readily that (\qnine) has the canonical form of equations associated with the affine Weyl group E$_7^{(1)}$. Thus, thanks to the chain of Miura transformations (\qtwo-\qthree) and (\qseven-\qeight), together with the translation (\qfive), we were able to bring an equation, associated with the group E$_6^{(1)}$ but written in the canonical form for equations coming from the group E$_8^{(1)}$, to a form which is canonical for equations of the group E$_7^{(1)}$. Whether one can proceed further and bring equation (\tten) to the canonical form for equations associated to the group E$_6^{(1)}$, i.e. an equation of the form $(X_n+X_{n+1})(X_n+X_{n-1})=f(X_n)$, remains an open question. 

It is interesting to point out here that the autonomous limit of equation (\qsix), or equivalently (\qfour), does not lead to a QRT-type mapping, but rather to a form first encountered in [\refdef\tsuda]. We find that the quantity
$$K(u_n,u_{n-1})={u_nu_{n-1}(u_nu_{n-1}-4z(u_n-u_{n-1})\over u_n-u_{n-1}-2z}\eqdef\qten$$
is conserved under the conservation law $K(u_n,u_{n-1})=-K(u_{n+1},u_n)$. This means that the quantity truly conserved by the autonomous limit of (\qsix) is the square of $K$, leading again to an HKY-type invariant.  

\bigskip
6. {\scap Conclusions}
\medskip

This paper was motivated by an exploratory calculation of possible integrable systems of a particular form. As explained in section 2, the form of the equations we set out to examine was suggested by that of the additive discrete Painlev\'e equations associated to the affine Weyl group E$_8^{(1)}$. Our calculation, based on the algebraic entropy method, resulted in two new (among five in all) integrable autonomous mappings, one of which was expected to be linearisable. Moreover, the two mappings in question were not of QRT type but rather of the class of systems [\hky]  discovered by Hirota, Kimura and Yahagi, and referred to as HKY. The invariant of an HKY mapping is not bi-quadratic as in the QRT case, but involves higher degrees (bi-quartic for the two cases under consideration). 

We performed the deautonomisation of these two new mappings. Given the limiting procedures that underlie their particular forms, as postulated in  section 2, we did not expect the non-autonomous versions of these mappings to be associated to the group E$_8^{(1)}$. In fact, one of the equations turned out to be related to the group E$_7^{(1)}$ while the other one was linearisable. For the first equation we introduced a series of Miura transformations which allowed us to bring the equation to the canonical form of discrete Painlev\'e equations associated to the group E$_7^{(1)}$. This chain of Miura transformations, when applied at the autonomous limit, also had the effect of transforming the HKY-type mapping to a QRT one. In the case of the linearisable mapping, the deautonomisation resulted into an equation involving one free function. Here again we introduced a chain of Miura transformations that allowed us to effectively linearise the equation and to transform the HKY-type invariant for that equation to a QRT one. 

We also examined a third equation among the five detected in section 2, obtained for $N=10$. Deautonomising it, we recognised an equation associated to the affine Weyl group E$_6^{(1)}$, already obtained in [\ourlimit]. Again we considered the possible Miura transformations and while the equation is indeed of QRT type at the autonomous limit, the first Miura cast it in a non-QRT form, reminiscent of systems we studied in [\tsuda]. A final Miura transformation brought the equation to a form that is canonical for discrete Painlev\'e equations associated to the group E$_7^{(1)}$ (and it is not clear whether a transformation to a canonical E$_6^{(1)}$ form is possible).

The main conclusion of this paper is that there exists an extreme richness in the Miura transformations relating various discrete Painlev\'e equations. Moreover, it is now clear that one can obtain new forms for the latter starting from HKY-type mappings, relying on the remarkable effect these Miura transformations have on the invariants for the associated autonomous systems, relating HKY and QRT-type invariants. Moreover, the same conclusions apply to linearisable mappings as well. While our approach was based on systems of additive type it is clear that it can be extended to systems of multiplicative type. We intend to return to this question in the future. 

\bigskip
{\scap Acknowledgements}
\medskip

RW would like to acknowledge support from the Japan Society for the Promotion of Science (JSPS), through the the JSPS grant: KAKENHI grant number 15K04893. 

\bigskip
{\scap References}
\medskip

 [\infinit] A. Ramani and B. Grammaticos, Phys. Lett. A 373 (2009) 3028.
 
 [\gromak] V.A. Gromak and N.A. Lukashevich, {\sl The analytic solutions of the
Painlev\'e equations}, (Universitetskoye Publishers, Minsk 1990), in Russian.

 [\fold] T. Tsuda , K. Okamoto and H. Sakai, Math. Ann. 331 (2005) 713.

 [\miura] R. Miura, J. Math. Phys. 9 (1968) 1201.

 [\malm] J. Malmquist, Arkiv. Math. Astr. Fys. 17 (1922) 1.
 
 [\bureau] F.J. Bureau, Annali di Matematica 64 (1964) 229.

 [\okamo] K. Okamoto, Physica D 2 (1981) 525.
 
 [\cosgr] C. Cosgrove and G. Scoufis, Stud. Appl. Math. 88 (1993) 25.

 [\ourmiura] A. Ramani and B. Grammaticos, J. Phys. A 25 (1992) L633.

 [\fokas] A. Fokas, B. Grammaticos and A. Ramani, J. of Math. Anal. and Appl. 180 (1993) 342.

 [\ourlimit] K.M. Tamizhmani, T. Tamizhmani, A. Ramani, B. Grammaticos, {\sl On the limits of discrete Painlev\'e equations associated to the affine Weyl group E$_8$}, preprint (2017).
 
 [\qrt] G.R.W. Quispel, J.A.G. Roberts and C.J. Thompson, Physica D34 (1989) 183.
 
 [\hky] R. Hirota, K. Kimura and H. Yahagi, J. Phys. A. 34 (2001) 10377.
 
 [\huit] Y. Ohta, A. Ramani and B. Grammaticos, J. Phys. A 34 (2001) 10523.

 [\ourdegen] A. Ramani and B. Grammaticos, J. Phys. A 50 (2017) 055204.
 
 [\kanoya] K. Kajiwara, M. Noumi and Y. Yamada, J. Phys. A 50 (2017) 073001.

 [\sincon] B. Grammaticos, A. Ramani and V. Papageorgiou, Phys. Rev. Lett. 67 (1991) 1825.

 [\seven] A. Ramani, R. Willox, B. Grammaticos, A.S. Carstea and J. Satsuma, Physica A 347 (2005) 1.

 [\tsuda] T. Tsuda, A. Ramani, B. Grammaticos and T. Takenawa, Lett. Math. Phys. 82 (2007) 39.

\end{document}